\documentstyle[11pt,newpasp,twoside,epsf]{article}
\def\edcomment#1{\iffalse\marginpar{\raggedright\sl#1\/}\else\relax\fi}
\reversemarginpar

\begin{document}
\title{V2487 Ophiuchi: A strong candidate for a recurrent nova
and a progenitor of Type Ia supernova
} 

\author{Izumi Hachisu} 

\affil{College of Arts and Sciences, University of Tokyo, 
Meguro-ku, Tokyo 153-8902, Japan}

\author{Mariko Kato} 

\affil{Keio University, Kouhoku-ku, Yokohama 223-8521, Japan}

\author{Taichi Kato and Katsura Matsumoto} 

\affil{Kyoto University, Sakyo-ku, Kyoto 606-8502, Japan}

\begin{abstract}
     The light curve of the 1998 outburst of V2487 Ophiuchi 
(Nova Oph 1998) shows a very rapid decline ($t_3 \sim 9$ day)
and a mid-plateau phase from 10 to 30 days after the optical maximum,
which are characteristics common to the U Sco subclass of the recurrent 
novae.  We have numerically reproduced light curves 
of the 1998 outburst based on a thermonuclear runaway model with 
optically thick winds.  The results show that
the mass of the white dwarf (WD) is as massive as $1.35 \pm 0.01 M_\odot$;
the envelope mass of the WD at the optical maximum is 
$\sim 6 \times 10^{-6} M_\odot$; the hydrogen content of 
the WD envelope is as low as $X \sim 0.1$ by mass weight.  
The mass transfer rate in quiescence is roughly estimated to be 
$\sim 1.5 \times 10^{-7} M_\odot$ yr$^{-1}$ and, therefore, 
that the recurrence period of nova outbursts is about 40 yr.
Since the WD mass may be now growing at a rate of 
$\sim 2 \times 10^{-8} M_\odot$ yr$^{-1}$,
V2487 Oph is a strong candidate for a progenitor of a Type Ia
supernova.
\end{abstract}

\section{Light Curve Analysis and a Very Massive White Dwarf}

     The outburst of V2487 Ophiuchi (Nova Oph 1998) was discovered 
at $m_V \sim 9.5$ by Takamizawa on 1998 June 15.561 UT.
The early optical decline rate of 0.37 mag day$^{-1}$ makes 
this nova one of the fastest ever seen 
(see Fig. 1, from the VSNET archives). 
The rapid decline stopped about 10 days after maximum and then
the decline rate lessened to $\sim 0.05$ mag day$^{-1}$, that is,
the brightness stayed at $m_V \sim 14$ from 10 to 30 days 
after maximum.  We call this period {\it mid-plateau phase}.  
Then the optical brightness again declined so fast from $m_V \sim 14$ 
to $m_V \sim 16$ in about 10 days.  The brightness stayed again 
at $m_V \sim 16$ at least until 70 days after maximum
(HJD 2,451,050), which is not shown in Fig.1 but from observational 
points reported in the AAVSO archives.  
\par
     We have modeled the system consisting of a very massive WD and
a lobe-filling main-sequence (MS) star.  Irradiation effects of 
the accretion disk (ACDK) and the MS companion by the WD are included 
into the light curve calculation.  The numerical method has been 
described in Hachisu \& Kato (2001).  We are able to reproduce 
the light curve by adopting model parameters similar to those 
of U Sco (Hachisu et al. 2000).  The model parameters 
are shown in Fig. 1.
\par
     The very rapid decline and the ensuing mid-plateau phases are 
common features among the U Sco subclass of recurrent novae.
If V2487 Oph belongs to the U Sco subclass, its orbital period 
should be between $\sim 0.3$ and $\sim 3$ days.
The numerical results strongly indicate that this nova is a recurrent 
novae.  Adding V2487 Oph to the member of the recurrent novae would
provide us valuable information on recurrent novae and SN Ia progenitors.

\begin{figure}
\plotone{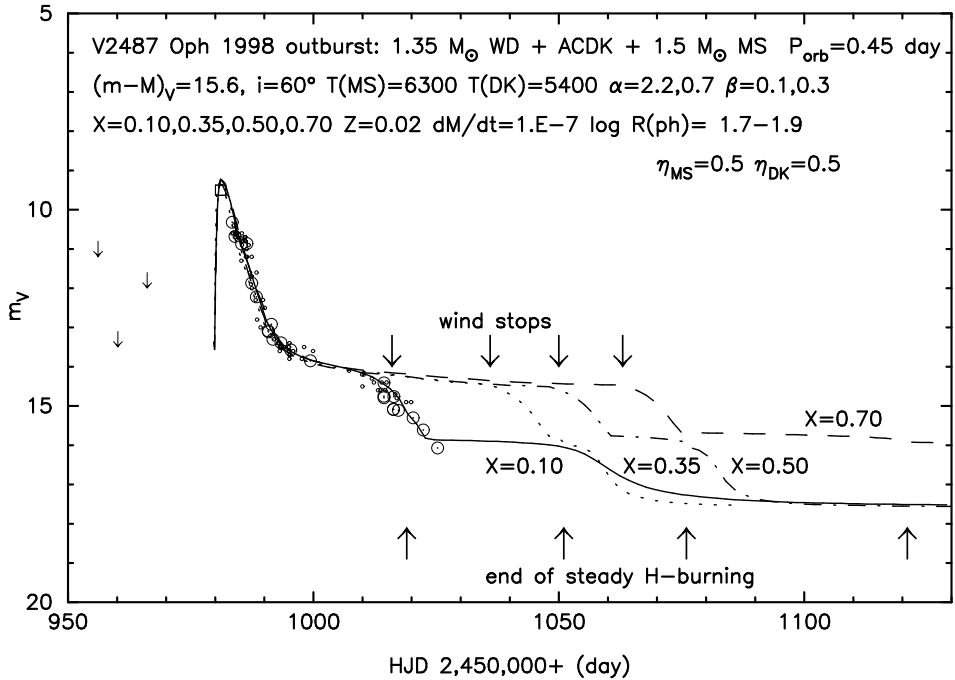}
\caption{
Calculated $V$ light curves are plotted against time (HJD 2,450,000+) 
together with the observational points of the 1998 outburst (all 
taken from the VSNET archives).
Each line indicates the light curve of $M_{\rm WD}= 1.35 M_\odot$ 
connecting the $V$ light at the binary phase 0.5 
(the phase at which the WD component is in front of the MS companion
seen from the Earth) for
various hydrogen contents of the WD envelope, i.e.,
$X=0.70$ (dashed line), $X=0.50$ (dash-dotted line), 
$X=0.35$ (dotted line), and $X=0.10$ (thick solid line).
}
\end{figure}

\end{document}